# Disturbance Observer-based Robust Control and Its Applications: 35th Anniversary Overview

Emre Sariyildiz, *Member, IEEE*, Roberto Oboe, Senior *Member, IEEE*, Kouhei Ohnishi, Life *Fellow, IEEE*

*Abstract*—**Disturbance Observer (DOb) has been one of the most widely used robust control tools since it was proposed by K. Ohnishi in 1983. This paper introduces the origins of DOb and presents a survey of the major results on DOb-based robust control in the last thirty-five years. Furthermore, it explains DOb's analysis and synthesis techniques for linear and nonlinear systems by using a unified framework. In the last section, this paper presents concluding remarks on DOb-based robust control and its engineering applications.**

*Index Terms* — **Disturbance Observer, Robust Control.**

## I. Introduction

WITH the severe sensitivity problem of the optimal control theory in engineering applications, robust control was emerged to deal with plant uncertainties and external disturbances in the beginning of 1970s. Several robust control techniques since then have been proposed in the literature. Among them, DOb is one of the most popular robust control tools due to its simplicity, flexibility and efficacy. In DOb-based robust control, internal and external disturbances are estimated by using identified dynamics and measurable states of plants, and the robustness of systems is simply achieved by feedbacking the estimations of disturbances. In the last four decades, this intuitive robust control technique has been experimentally verified in many different engineering applications, such as in robotics, mechatronics, automotive and power electronics. The objective of this paper is to provide an overview on DOb-based robust control and its engineering applications. In this paper, DOb-based robust control technique is exemplified in the motion control framework.

The paper is organized as follows. In section II, the origins of DOb is introduced. In section III, the major results on DOb-based robust control and its applications are presented. In section IV, DOb is synthesized in frequency and time domains. In section V, a Two-Degrees-of-Freedom (2-DoF) robust control system is synthesized by using DOb. In section VI, the paper ends with conclusion and remarks.

## II. Origins of Disturbance Observer

### A. Birth of Robust Control (1960s – 1980s):

Plant uncertainties and external disturbances are inevitable and indispensable in many engineering systems; e.g., robots, hard-disk drives, chemical reactors and spacecraft [1]–[5]. Feedback controllers are designed so that the performance goals of systems can be achieved by attenuating disturbances in real implementations. Classical feedback control methods, such as Bode and root-locus, implicitly synthesize robust controllers with limited disturbance suppression capability [6, 7]. Horowitz, for the first time, analytically formulated the trade-off between the robustness and performance of classical feedback control systems without explicitly using the robustness term in 1963 [8]. However, the significance of Horowitz's contribution was not recognized in 1960s as the large model-plant mismatches were generally neglected in the *ad hoc* methods of classical design [7, 8]. At the same time, modern optimal control theory received increasing attention to tackle more complex control problems such as control of nonlinear and Multi-Input-Multi-Output (MIMO) systems [9]. Although modern control techniques provide strong mathematical tools, as well as wider application area, they are more sensitive to disturbances than classical control techniques. In the early 1970s, many researchers reported the failures of modern optimal control techniques due to lack of robustness [9]–[11]. A consensus was immediately reached on the importance of treating large disturbances in the design of controllers; and the robust control field was born in 1970s [9]. To describe the tracking performance of a system which suffers from disturbances, the robustness term was first explicitly used by Pearson and Stats in 1974 and Davison in 1975 [12, 13]. In the following years, many robust control techniques were proposed to improve the stability and performance (i.e., robust stability and robust performance) of control systems in the presence of plant uncertainties and external disturbances; e.g., H∞ control, Sliding Mode Control (SMC), Structured Singular Values (SSV) or µ-synthesis, Internal Model Control (IMC) and Robust Parametric Control

E. Sariyildiz is with the School of Mechanical, Materials, Mechatronic and Biomedical Engineering, University of Wollongong, Wollongong, NSW 2522, Australia (corresponding author: phone: +61-242213319; fax: +61- 242214577; e-mail: emre@uow.edu.au).

such as Kharitonov's theorem [14]–[18]. The failures of modern optimal control techniques have resulted in a significant paradigm shift from optimality to robustness in control theory. Today, robustness against disturbances, as well as achieving stability and performance goals, has become the key objective of a feedback controller synthesis.

In general, robust control techniques can be divided into two categories: suppressing disturbances via feedback control, such as IMC and SSV, and cancelling disturbances via feedforward control [16, 17, 19]. The former has been well-developed in the last four decades. However, it generally synthesizes complex controllers, it cannot react fast enough in the presence of strong disturbances although they can be eventually suppressed, and conservatism is a challenging problem in many robust feedback control methods such as H∞ control [20, 21]. In the latter, the reverse of disturbance signal is feedforwarded so that the robustness of a system is intuitively achieved by cancelling disturbances [19, 20]. The main drawback of this robust control technique is that disturbances are unknown and unmeasurable in many engineering systems so the robust feedforward controller synthesis is generally impractical. To tackle this problem, many observers have been proposed to estimate disturbances by using the measurable states and known dynamics of plants. The robustness of a system is similarly achieved by feedforwarding the estimations of disturbances instead of exact disturbances. In fact, a robust feedback controller is implicitly synthesized due to the dynamics of disturbance estimation when observers are used in the robust feedforward control technique. Therefore, observer-based robust controllers have also been described as an IMC method by many researchers in the literature [22, 23].

### B. Birth of DOb (1960s – 1980s):

DOb is the most popular robust control tool that is used to estimate plant uncertainties and external disturbances [24]. Similar to the robust control theory, the origins of DOb can be traced back to the 1960s. To deal with the sensitivity problems of conventional state observers (e.g., Luenberger observer), robust state observers were proposed by considering unknown/unmeasured inputs (i.e., external disturbances) of a system in 1960s and 1970s [25, 26]. Soon, it has been noticed that an Unknown Input Observer (UIO) that estimates external disturbances can be designed by slightly modifying robust state observers [27]. For the first time, Johnson improved the robustness of a control system by implicitly using the estimations of constant disturbances in 1968 and general disturbances in 1970 [28, 29]. In the following year, he designed an optimal robust controller, namely Disturbance Accommodating Controller (DAC), by explicitly using the estimations of external disturbances [30]. It is worth noting that Johnson made very important contributions to the observer-based robust control theory in the 1970s. For example, in 1973, he designed a robust controller by using an observer that estimates not only the external disturbances but also the states of a system [31]. Today, this robust control tool is known as Extended State Observer (ESO) in the literature [19]. The model-plant mismatches were neglected in the conventional DAC synthesis. In 1985, an adaptive DAC was proposed to deal with plant uncertainties in addition to external disturbances [32]. However, the practical significance of the observer-based robust controller synthesis was not noticed due to the complex optimal robust control structure of DAC. Ohnishi, for the first time, proposed the DOb to estimate the external disturbances of a servo system by using Gopinath's reduced-order observer design method (aka auxiliary variable-based observer design method) in 1983 [33]. Similar to DAC, an optimal controller with an integrator was implemented so that optimal performance was achieved by suppressing plant uncertainties while external disturbances were cancelled/suppressed with their estimations via DOb. With the auxiliary variable-based observer design method, the dynamics of disturbance estimation was explicitly formulized as a low-pass filter (LPF); i.e., the bandwidth of disturbance estimation was clearly described. After the first paper of DOb, an implicit DOb-based robust motion control system was proposed by Ohnishi in 1985 [34]. This work has clarified the 2-DoF control structure of a DOb-based robust control system by using classical control techniques. It was theoretically and experimentally proven that the robustness and performance of a control system can be independently adjusted by using a DOb and a performance controller (e.g., a PD controller), respectively. To deal not only with the external disturbances but also with the plant uncertainties of a system, Ohnishi explicitly synthesized a DOb-based 2-DoF robust controller in 1987 [35]. By lumping nonlinear plant dynamics and internal and external disturbances together into a fictitious disturbance variable, a linear nominal plant model was used in the design of DOb. In other words, a nonlinear system was linearized without requiring the precise dynamic model of the plant, and a robust controller was practically synthesized for a nonlinear system by using linear control methods. The DOb-based 2-DoF robust controller was experimentally verified by performing the decentralized position control of a robot manipulator in this paper. In the 1980s, Ohnishi elegantly and cogently explained the practical significance of the observer-based robust control by shifting its analysis and synthesis techniques from time domain to frequency domain. His proposals have been applied to many complex engineering systems in the last thirty-five years [36–39].

### III. DEVELOPMENT OF DOb-BASED ROBUST CONTROL

### A. DOb-based Robust Control (1990s):

Observer-based robust control received increasing attention by control practitioners in 1990s. DOb was applied to many different engineering applications from motion control of servo systems and CNC machines to power electronics, system identification and fault diagnosis [40] – [44]. Loop-shaping control techniques, such as Bode/Nyquist plots and H∞ control, were generally used in continuous and discrete time domains in order to tune the stability and performance of the robust controller, i.e., the outer-loop performance controller, nominal plant dynamics and LPF of DOb. For example, Hori tuned the robustness and noise-sensitivity by shaping the frequency responses of the sensitivity and complementary sensitivity functions, respectively, in [40], Tomizuka improved the tracking performance of a servo system by combining Zero Phase Error Tracking Controller (ZPETC) and DOb in [45], Kempf showed that the design constraints of DOb change when the plant includes time-delay in [46] and Mita proposed a new robust H∞ controller by using DOb in [47]. Although they were not as popular as the loop-shaping control methods, there were also few examples of advanced DOb-based robust control methods in 1990s; e.g., nonlinear analysis and synthesis of the observer and robust controller [48] – [50], robust model predictive controller synthesis [51, 52], suppressing the chattering of an SMC controller by eliminating disturbances [53], robust repetitive learning control [54], robust fuzzy logic control [55], and the optimal robust controller synthesis by using Linear Quadratic Integral method [56]. In addition, there were some important theoretical results on the observer-based robust control method; e.g., the necessary and sufficient conditions for the existence of an observer with unknown inputs were given in [57, 58] and the equivalence of the passivity and DOb based robust control systems was shown in [48]. However, the significance of these studies was not recognized among control practitioners in 1990s due to complex control structures, practical limitations in implementations and lack of analysis and synthesis tools. Active Disturbance Rejection (ADR) control proposed by Han started to become popular in these years, particularly in China because it was originally published in Chinese [19]. An ADR controller is conventionally synthesized by using an ESO and minimum information for the nominal plant model (i.e., only the relative degree of the plant). Synthesizing the robust controller by using a simple dynamic model for complex systems is one of the most important advantageous of the DOb-based robust control technique (e.g., Ohnishi used a linear nominal plant model for a robot manipulator in [35]). Although a simple control law can be theoretically obtained by ignoring the complex dynamics of plants, this oversimplification has severe limitations in practice. The stability and performance of an observer-based robust control system may significantly deteriorate if the plant-model mismatches cannot be compensated due to practical design constraints such as noise and sampling time [23, 59]. The conventional ADR controller has several limitations in practice [59] – [62]. However, Han's philosophical discussion on PID and ADR control frameworks has inspired many researchers in the last two decades [19].

### B. DOb-based Robust Control (2000s):

In addition to servomotors, robots and power electronics, DOb was applied to many different engineering systems, such as automobiles, electric commuter trains, networks, missile seekers, and spacecraft, in the last two decades [63] – [67]. High performance engineering applications have motivated not only control practitioners but also control theoreticians to study DOb-based robust control method. Today, if "disturbance observer" is searched in *IEEE Xplore*, then more than 990 Journals & Magazines and 3970 conference papers published between 2010 and 2019 are found. The same search results with 282 Journals & Magazines and 1309 conference papers published between 2000 and 2009 and 100 Journals & Magazines and 445 conference papers published between 1990 and 1999. More and more researchers are adopting the DOb-based robust control method every year.

In order to improve the stability and performance of DOb-based robust control applications, more rigorous analysis and synthesis techniques were proposed by using linear control methods in 2000s. Compared to 1990s, not only the dynamics of the LPF of DOb, e.g., the bandwidth and order of the LPF, but also the nominal plant model and outer loop controller were thoroughly considered in the analysis. It was shown that the robust stability and performance significantly change when the plant includes right-half-plane pole(s)/zero(s) and/or time-delay [23]. For example, an almost necessary and sufficient condition for the robust stability of the controller was given for minimum phase systems in [68], the exact condition was derived for minimum phase plants with parametric uncertainties in [69], a guide for the DOb-based robust controller synthesis was proposed for minimum and non-minimum phase systems by using Bode Integral Theorem in [21, 23], a generalized DOb was proposed to estimate higher-order disturbances in [70], a sensitivity optimization approach was proposed for digital implementation in [71], periodic disturbances were suppressed by using a periodic DOb in [72], the stability of the robust controller was analyzed under time delay in [73], and frequency and time domain analysis and synthesis techniques were developed in [74, 75]. The proposed linear control methods were widely adopted and applied to many different engineering applications by control practitioners. Today, we have several advanced linear control tools to analyze and synthesize DOb-based robust control systems.

On the other hand, nonlinear control techniques became popular in the analysis and synthesis of DOb-based robust control systems in the early 2000s. X. Chen and Fukuda proposed a nonlinear DOb synthesis technique by using variable structure system theory in 2000 [76]. In the same year, W. Chen showed that the performance of disturbance estimation –thus the robustness of the control system– can be improved by using a nonlinear nominal dynamic model in the auxiliary variable-based design method [62]. The stabilities of the DOb and robust controller were proved by using the Lyapunov's direct method [77]. When more complex nominal dynamic models (e.g., nonlinear dynamics of plants) are used in the design of DOb: better disturbance estimation can be achieved, the stability and performance of the robust control system can be improved, the bandwidth of DOb can be increased in practice and the trade-off between the robustness and noise-sensitivity can be tuned better [78]. Moreover, the nonlinear analysis and synthesis techniques help extend the application areas of DOb. The robust controller synthesis, however, may become less intuitive. Several nonlinear analysis and synthesis techniques since then have been developed for DOb-based robust control systems. For example, DOb was synthesized for multivariable nonlinear systems in [79], the robust controller was synthesized in a back-stepping manner and an input-to-state nonlinear stability analysis was proposed in [80], a linear DOb was extended to nonlinear systems in [81], a modular design method was proposed in [82], and a nonlinear stability analysis was proposed by considering the practical design constraints of DOb in [78]. With the proposed analysis and synthesis techniques, DOb was applied to different nonlinear systems such as a Duffing–Holmes chaotic system and nonlinear multi-agent systems [83, 84].

The flexible structure of the 2-DoF control method has allowed researchers to develop different DOb-based robust controllers in the last two decades. Several controllers were synthesized by combining DOb and an advanced control method, such as intelligent control, so that the robust stability and performance of the latter were improved. Some examples of these controllers in 2018 are as follows: a Neural Network controller was combined with an observer to control the trajectory of an underwater vehicle in [85], a Model Predictive controller was combined with an observer to control three-phase inverters in [86], an SMC controller was combined with an observer to control fractional order systems in [87] and H∞ and resilient control methods were combined with an observer to control a nonlinear singular stochastic hybrid system with partly unknown Markovian jump parameters in [88]. The recent trend in DOb research shows that we will see more examples of the advanced 2-DoF robust controllers in the future. Nevertheless, the analysis and synthesis of the advanced robust controllers are more complicated than that of the conventional DOb-based robust controllers.

### C. Observer-based Robust Control and Applications:

Robust motion control (e.g., position, force and compliance control) has been one of the most popular application areas of DOb since 1980s. The 2-DoF robust controller was applied to the motion control problem of many different engineering systems, spanning from industrial robots, hard-disk drives, automobiles and Hubble Space Telescope to today's cutting edge robotic systems such as compliant exoskeletons, surgery robots and unmanned aerial vehicles [35] – [37], [89-92]. It was shown by Murakami that a DOb can be used not only to achieve the robustness of a motion control system but also to estimate contact force/torque as a force sensor [93]. This specific application of DOb is known as Reaction Force Observer (RFOb) in the literature, and it has been verified in many different engineering applications, e.g., in rehabilitation robotics, electric commuter trains and automotive [38, 64, 89]. As a robust motion control tool, DOb has been rapidly matured in the last three decades. Today, DOb-based motion control products are commercially available in the market, and new motion control products keep being developed. For example, DOb was embedded in the Panasonic's MINAS-A5 series motor drivers to diminish the impact of the disturbance torque, reduce vibration, and offset any speed decline [94]; and a new LSI module was developed for real haptics by using DOb and RFOb in [95, 96].

In addition to motion control, DOb has been applied to several applications in different research fields in the last thirty-five years. For example, robotic eye-hand calibration [97], robust control of mineral grinding process [98], time delay estimation and compensation of network control systems [99], DC-bus voltage control of micro-grid systems [100], deadbeat control for UPS [101], partial synchronization of neurons [102], sensorless measurement of pulsatile flow rate [103], temperature control of a superheated steam [104], and feedback linearization control of a nuclear reactor [105]. Since plant dynamics are more complicated than servo systems, nonlinear nominal plant models and advanced performance controllers, such as nonlinear and iterative learning controllers, have been generally employed in the design of the

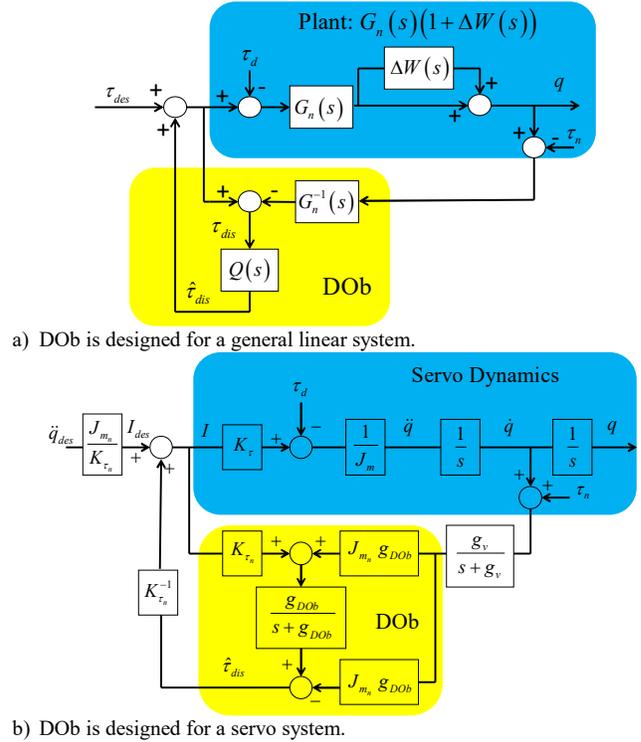

a) DOb is designed for a general linear system.

b) DOb is designed for a servo system.

Fig.1: Block diagrams of DOb.

robust controllers [87, 106, 107]. Although the performance controller and DOb can be independently synthesized thanks to the flexible structure of the 2-DoF control method, the analysis and synthesis of the advanced robust controllers are generally complicated.

In the last three decades, DOb-based robust control has inspired many researchers. Several linear and nonlinear observer-based robust controllers have been independently developed in the literature. For example, the robust controllers based on: Perturbation Observer (PO), Equivalent Input Disturbance (EID) estimator, Uncertainty and Disturbance Estimator (UDE), Generalized Proportional Integral Observer (GPIO), Extended State Observer (ESO) and Extended High Gain Observer (EHGO) [108] – [112]. In these robust control systems, the fundamental idea behind the controller synthesis is same as the DOb-based robust control system: estimating the internal and external disturbances by using the known dynamics and measurable states of a system (i.e., disturbance observer synthesis), feedbacking the estimations of disturbances so as to intuitively achieve the robustness of the control system (i.e., disturbance cancellation/suppression) and tuning the performance controller by considering the nominal plant model (i.e., 2-DoF control). Therefore, they are described as DOb-based robust control in this paper. The reader is recommended to refer to [113] for a recent comprehensive survey on the observer-based robust control techniques.

## IV. ANALYSIS AND SYNTHESIS OF DOB

### A. Classical Control:

Block diagram of a DOb is illustrated in Fig. 1a. In this figure, $G_n(s)(1+\Delta W(s))$ represents plant dynamics with unstructured uncertainty in which $G_n(s)$ is nominal plant model, $W(s)$ is a fixed stable transfer function, the weight, and $\Delta(s)$ is a variable stable transfer function satisfying $\|\Delta(s)\|_\infty \leq 1$; $\tau_d$ and $\tau_n$ represent exogenous disturbance and noise inputs, respectively; $q$ represents output; $\tau_{des}$ represents the control signal of the performance controller; $Q(s)$ represents the LPF of DOb; $s$ represents Laplace variable; $\tau_{dis}$ represents a fictitious disturbance variable which includes internal and external disturbances; and $\hat{\tau}_{dis}$ represents the estimation of $\tau_{dis}$.

By neglecting the exogenous noise input, the relation between input and output is derived from Fig. 1a as follows:

$$q = G_n(s)(\tau_{des} + \hat{\tau}_{dis} - \tau_{dis}) \quad (1)$$

where $\tau_{dis} = (1+\Delta W(s))\tau_d - \Delta W(s)(\tau_{des} + \hat{\tau}_{dis})$ and $\hat{\tau}_{dis} = Q(s)\tau_{dis}$.

Eq. (1) shows that if the bandwidth of DOb is large enough, (i.e., $Q(s) \cong 1$ and $\hat{\tau}_{dis} \cong \tau_{dis}$), then the plant uncertainties and external disturbances are precisely eliminated with their estimations, and thus the performance controller can be designed by considering only the nominal plant dynamics. Indeed, this simple analysis works very-well in many practical applications such as robust motion control of robot manipulators. This is why frequency domain analysis and synthesis techniques of DOb have been widely adopted by control practitioners.

Since the dynamics of the LPF of DOb directly influences the estimations of disturbances, its synthesis has received special consideration by many researchers [23, 34, 40]. The performance of disturbance estimation, and thus the robustness of the control system (See Fig. 1a), can be simply improved by either increasing the bandwidth of DOb or using a higher order LPF [23]. However, the bandwidth and the order of the LPF of a DOb are limited by practical and theoretical design constraints, e.g., noise and the waterbed effect, respectively [23, 42]. The former depends on the specifications of control equipment such as resolution of an encoder and sampling time of a real-time controller, and the latter depends on the dynamics of plants, such as RHP zeros and poles [23, 42].

The sensitivity and complementary sensitivity transfer functions of a DOb-based robust control system are derived from Fig. 1a as follows:

*Sensitivity Function:*

$$S_{DOb} = \frac{1-Q(s)}{1-Q(s)+(1+\Delta W(s))Q(s)} \quad (2)$$

*Complementary Sensitivity Function:*

$$T_{DOb} = 1 - S_{DOb} = \frac{(1+\Delta W(s))Q(s)}{1-Q(s)+(1+\Delta W(s))Q(s)} \quad (3)$$

where $S_{DOb} = 1-Q(s)$ and $T_{DOb} = Q(s)$ when $W(s) = 0$, i.e., plant model is precisely known; $S_{DOb} \to 0$ and $T_{DOb} \to 1$ as $s = j\omega \to 0$, i.e., at low frequencies; and $S_{DOb} \to 1$ and $T_{DOb} \to 0$ as $s = j\omega \to \infty$, i.e., at high frequencies.

Eq. (2) and Eq. (3) show that a DOb-based robust controller can precisely suppress disturbances and noise at the asymptotic frequencies. However, the dynamic responses of the sensitivity and complementary sensitivity functions depend on the plant uncertainties and the LPF of DOb at the middle frequencies. In other words, the robustness, stability and performance of a DOb-based control system can be adjusted by tuning either the bandwidth of DOb or the nominal plant dynamics.

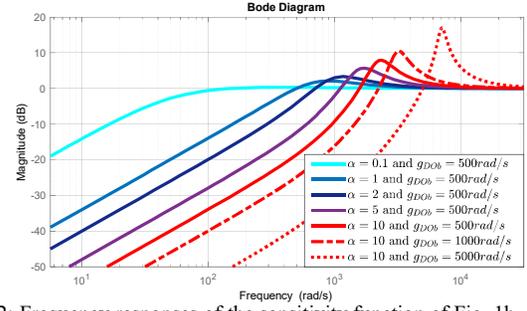

Fig.2: Frequency responses of the sensitivity function of Fig. 1b, i.e., Eq. (4), when $g_v = 1000\,\text{rad/s}$.

To show how the bandwidth of LPF and the dynamics of nominal plant model influence the robust stability and performance, let us consider a DOb-based robust motion control system which is illustrated Fig. 1b. In this figure, $J_m$ and $J_{m_n}$ represent the uncertain and nominal inertias, respectively; $K_\tau$ and $K_{\tau_n}$ represent the uncertain and nominal thrust coefficients, respectively; $q, \dot{q}$ and $\ddot{q}$ represent the angle, velocity and acceleration of the servo system, respectively; $g_{DOb}$ and $g_v$ represent the bandwidths of the LPF of DOb and velocity measurement, respectively; $I$ represents the motor current; and $I_{des}$ and $\ddot{q}_{des}$ represent the desired $I$ and $\ddot{q}$, respectively.

The sensitivity function of a DOb-based robust motion control system is derived from Fig. 1b as follows:

$$S_{DOb} = \frac{s(s+g_v)}{s^2 + g_v s + \alpha g_v g_{DOb}} \quad (4)$$

where $\alpha = (J_{m_n} K_\tau)/(J_m K_{\tau_n})$.

Eq. (4) and Fig. 2 show that the robustness against disturbances can be simply improved at low frequencies by increasing either the bandwidth of DOb or $\alpha$. Higher values of $\alpha$ can be obtained by increasing $J_{m_n}$ or decreasing $K_{\tau_n}$. However, the robust motion control system becomes more noise sensitive (See the waterbed effect in Fig. 2) [23]. The robust performance and stability of the motion control system deteriorate as the peaks of the sensitivity and complementary sensitivity transfer functions are increased at the middle frequencies as shown in Fig. 2.

### B. Modern Control:

To synthesize a DOb in state space, let us consider the following dynamic model.

$$\begin{aligned}\dot{\mathbf{x}} &= \mathbf{A}\mathbf{x} + \mathbf{b}u - \boldsymbol{\tau}_\mathbf{d} \\ \dot{\mathbf{x}} &= \mathbf{A_n}\mathbf{x} + \mathbf{b_n}u - \boldsymbol{\tau}_\mathbf{dis}\end{aligned} \quad (5)$$

where $\mathbf{x}$ and $\dot{\mathbf{x}} \in \mathbb{R}^n$ represent the state vector of the system and its time derivative, respectively; $\mathbf{A}$ and $\mathbf{A_n} \in \mathbb{R}^{n\times n}$ represent the uncertain and nominal system matrices, respectively; $\mathbf{b}$ and $\mathbf{b_n} \in \mathbb{R}^n$ represent the uncertain and nominal control input vectors, respectively; $u \in \mathbb{R}$ represents the control input; $\tau_d \in \mathbb{R}^n$ represents a disturbance vector which includes unknown plant dynamics and external disturbances; and $\tau_{dis} \in \mathbb{R}^n$ represents a disturbance vector which includes parametric uncertainties and $\tau_d$, i.e., $\tau_{dis} = (A_n - A)x + (b_n - b)u + \tau_d \in \mathbb{R}^n$. For the sake of simplicity, the dynamic model of a single input system is used in Eq. (5).

The disturbance vector $(\tau_{dis})$ can be estimated by designing a minimum order observer for the following augmented state space model of the system [24, 30].

$$\begin{bmatrix} \dot{x} \\ \dot{x}_{\tau_{dis}} \end{bmatrix} = \begin{bmatrix} A_n & -C_{\tau_{dis}} \\ 0 & A_{\tau_{dis}} \end{bmatrix} \begin{bmatrix} x \\ x_{\tau_{dis}} \end{bmatrix} + \begin{bmatrix} b_n \\ 0 \end{bmatrix} u \quad (6)$$

where $A_{\tau_{dis}} \in \mathbb{R}^{m \times m}$ and $C_{\tau_{dis}} \in \mathbb{R}^{n \times m}$ represent the system and output matrices of the disturbance model, i.e., $\dot{x}_{\tau_{dis}} = A_{\tau_{dis}} x_{\tau_{dis}}$ and $\tau_{dis} = C_{\tau_{dis}} x_{\tau_{dis}}$, respectively.

Eq. (6) shows that the dynamic model of disturbances should be known *a priori* in the design of DOb. This assumption is indeed very strict as the dynamics of disturbances is generally unknown in engineering systems. A DOb, however, can precisely estimate disturbances by using very simple dynamic models; e.g., a constant disturbance model $(\dot{\tau}_{dis} = 0)$ is generally used to estimate not only constant but also variable disturbances in practice. It has been theoretically and experimentally verified in many studies since DOb was proposed in 1983 [23, 114]. For example, Fig. 2 shows how variable disturbances are suppressed within the bandwidth of DOb. Of course, the performance of disturbance estimation can be improved with better approximation of the disturbance model; e.g., periodic disturbances are modeled to improve the robustness of DOb in [72]. Disturbance estimation can also be improved by using Generalized DOb [70].

Instead of the conventional minimum order observer-based design method, the state space synthesis of DOb has been generally performed by using the auxiliary variable design method due to its simplicity [24, 35].

### C. First-order DOb Synthesis using an Auxiliary Variable:

The estimation of the disturbance vector $(\tau_{dis})$ is derived in terms of an auxiliary variable and the states of the system as follows:

$$\hat{\tau}_{dis} = z - Lx \quad (7)$$

where $\hat{\tau}_{dis} \in \mathbb{R}^n$ represents the estimation of $\tau_{dis}$; $L \in \mathbb{R}$ represents the observer gain of DOb to be tuned; and $z \in \mathbb{R}^n$ represents the auxiliary variable vector which is derived by integrating

$$\dot{z} = L(A_n x + b_n u - \hat{\tau}_{dis}) \quad (8)$$

The derivative of Eq. (7) is obtained by using Eq. (8) as follows:

$$\dot{\hat{\tau}}_{dis} = L(A_n x + b_n u - \hat{\tau}_{dis}) - L(A_n x + b_n u - \tau_{dis}) \quad (9)$$

If $\dot{\tau}_{dis}$ is subtracted from both sides of Eq. (9), then

$$\dot{e}_{\tau_{dis}} = -L e_{\tau_{dis}} - \dot{\tau}_{dis} \quad (10)$$

where $e_{\tau_{dis}} = \hat{\tau}_{dis} - \tau_{dis} \in \mathbb{R}^n$.

Eq. (10) shows that asymptotic stability is achieved if $L$ is strictly positive and $\dot{\tau}_{dis} = 0$. As it is discussed in the minimum order observer-based design method, the latter assumption of the asymptotic stability is very strict. If a more practical assumption is made by using $\|\tau_{dis}\| \le \delta_{\tau_{dis}} \in \mathbb{R}$ and $\|\dot{\tau}_{dis}\| \le \delta_{\dot{\tau}_{dis}} \in \mathbb{R}$, then it can be shown that the error of disturbance estimation is uniformly ultimately bounded when $L$ is strictly positive, i.e.,

$$\|e_{\tau_{dis}}(t)\| \le \lambda \exp(-L(t - t_0)) \|e_{\tau_{dis}}(t_0)\| + \frac{\lambda}{L} \delta_{\dot{\tau}_{dis}} \quad (11)$$

where $\lambda > 0 \in \mathbb{R}$.

Eq. (11) shows that the convergence rate and the accuracy of disturbance estimation can be simply improved by increasing the observer gain $L$, i.e., the bandwidth of DOb. However, the asymptotic stability of disturbance estimation cannot be achieved when $\dot{\tau}_{dis} \ne 0$.

### D. High-order DOb Synthesis using Auxiliary Variables:

Not only disturbances but also their successive time derivatives can be estimated by using a higher order DOb. To estimate the disturbance vector and its successive time derivatives up to the order of k-1, a $k^{th}$ order DOb can be similarly designed as follows:

$$\begin{aligned} \hat{\tau}_{dis} &= z_1 - L_1 x \\ \dot{\hat{\tau}}_{dis} &= z_2 - L_2 x \\ &\vdots \\ \widehat{\tau}_{dis}^{(k-1)} &= z_k - L_k x \end{aligned} \quad (12)$$

where $\hat{\tau}_{dis} \in \mathbb{R}^n$ represents the estimation of the disturbance vector $\tau_{dis}$; $\dot{\hat{\tau}}_{dis}, \ddot{\hat{\tau}}_{dis}, \cdots, \widehat{\tau}_{dis}^{(k-1)} \in \mathbb{R}^n$ represent the estimations of the disturbance vector's successive time derivatives, i.e., $\dot{\tau}_{dis}, \ddot{\tau}_{dis}, \cdots, \tau_{dis}^{(k-1)} \in \mathbb{R}^n$, respectively; $L_j \in \mathbb{R}$ represents the $j^{th}$ observer gain to be tuned; and $z_j \in \mathbb{R}^n$ represents the $j^{th}$ auxiliary variable vector. The auxiliary variable vectors are derived by integrating

$$\begin{aligned} \dot{z}_1 &= L_1(A_n x + b_n u - \hat{\tau}_{dis}) + \dot{\hat{\tau}}_{dis} \\ \dot{z}_2 &= L_2(A_n x + b_n u - \hat{\tau}_{dis}) + \ddot{\hat{\tau}}_{dis} \\ &\vdots \\ \dot{z}_k &= L_k(A_n x + b_n u - \hat{\tau}_{dis}) \end{aligned} \quad (13)$$

where $\dot{z}_j \in \mathbb{R}^n$ represents the derivative of $z_j$.

Similar to the first order DOb analysis, if the $k^{th}$ order derivative of the disturbance vector is zero, then asymptotic stability can be achieved. However, if it is not zero but the disturbance vector and its successive time derivatives are bounded, then uniformly ultimately bounded estimation error can be achieved [20, 75]. The stability and performance of disturbance estimation are similarly adjusted by tuning observer gains as follows [20, 75]:

$$(\lambda + g_{DOb})^k = \lambda^k + \lambda^{k-1} L_1 + \lambda^{k-2} L_2 + \cdots + \lambda L_{k-1} + L_k \quad (14)$$

where $g_{DOb}$ represents the bandwidth of the $k^{th}$ order DOb.

The performance of disturbance estimation can be improved by using a higher order DOb. However, it is more noise-sensitive as it estimates the derivatives of disturbances; i.e., the bandwidth limitation becomes stricter as the order of DOb is increased. The reader is recommended to refer to [20, 23, 75] for further details on the higher-order DOb analysis and synthesis.

### E. DOb Synthesis using Nonlinear Dynamics:

A DOb can be similarly synthesized for a nonlinear system by using auxiliary variable design method. Let us consider the following nonlinear dynamic model to estimate the disturbance vector.

$$\dot{\mathbf{x}} = \mathbf{f}(\mathbf{x}) + \mathbf{g}(\mathbf{x})u - \boldsymbol{\tau}_{\mathbf{d}} \\ \dot{\mathbf{x}} = \mathbf{f_n}(\mathbf{x}) + \mathbf{g_n}(\mathbf{x})u - \boldsymbol{\tau}_{\mathbf{dis}} \quad (15)$$

where $\mathbf{f}(\mathbf{x})$ and $\mathbf{f_n}(\mathbf{x}) \in \mathbb{R}^n$ represent the nonlinear uncertain and nominal system vectors, respectively; and $\mathbf{g}(\mathbf{x})$ and $\mathbf{g_n}(\mathbf{x}) \in \mathbb{R}^n$ represent the nonlinear uncertain and nominal control input vectors, respectively. The other parameters are same as defined earlier; however, $\boldsymbol{\tau}_{\mathbf{dis}} = \boldsymbol{\tau}_\mathbf{d} + \mathbf{f_n}(\mathbf{x}) - \mathbf{f}(\mathbf{x}) + (\mathbf{g_n}(\mathbf{x}) - \mathbf{g}(\mathbf{x}))u \in \mathbb{R}^n$ in Eq. (15).

The estimation of the disturbance vector is derived in terms of an auxiliary variable and the states of the system as follows:

$$\hat{\boldsymbol{\tau}}_{\mathbf{dis}} = \mathbf{z} - \mathbf{L}(\mathbf{x}) \quad (16)$$

where $\mathbf{L}(\mathbf{x}) \in \mathbb{R}^n$ represents the observer gain vector to be tuned; and $\mathbf{z}$ is derived by integrating

$$\dot{\mathbf{z}} = \frac{\partial \mathbf{L}(\mathbf{x})}{\partial \mathbf{x}}\left(\mathbf{f_n}(\mathbf{x}) + \mathbf{g_n}(\mathbf{x})u - \hat{\boldsymbol{\tau}}_{\mathbf{dis}}\right) \quad (17)$$

The dynamic equation of disturbance estimation is derived by taking the derivative of Eq. (16) as follows:

$$\dot{\mathbf{e}}_{\boldsymbol{\tau}_{\mathbf{dis}}} = -\frac{\partial \mathbf{L}(\mathbf{x})}{\partial \mathbf{x}}\mathbf{e}_{\boldsymbol{\tau}_{\mathbf{dis}}} - \dot{\boldsymbol{\tau}}_{\mathbf{dis}} \quad (18)$$

where $\mathbf{e}_{\boldsymbol{\tau}_{\mathbf{dis}}} = \hat{\boldsymbol{\tau}}_{\mathbf{dis}} - \boldsymbol{\tau}_{\mathbf{dis}} \in \mathbb{R}^n$; and $\partial \mathbf{L}(\mathbf{x})/\partial \mathbf{x} \in \mathbb{R}^{n \times n}$ represents the observer gain matrix to be tuned.

Similar to Eq. (10), the stability of disturbance estimation can be achieved if $\partial \mathbf{L}(\mathbf{x})/\partial \mathbf{x}$ is a positive definite matrix. To adjust the stability and performance of disturbance estimation, a positive definite observer gain matrix, which satisfies Eq. (16) and Eq. (17), can be designed in different ways [62, 78].

Let us synthesize DOb for robot manipulators by intuitively tuning the observer gain matrix. The following nonlinear plant dynamics is used in the design of DOb.

$$\mathbf{M_n}(\mathbf{q})\ddot{\mathbf{q}} + \mathbf{C_n}(\mathbf{q},\dot{\mathbf{q}})\dot{\mathbf{q}} + \mathbf{g_n}(\mathbf{q}) = \boldsymbol{\tau} - \boldsymbol{\tau}_{\mathbf{dis}} \quad (19)$$

where $\mathbf{M_n}(\mathbf{q}) \in \mathbb{R}^{n \times n}$ represents the positive definite nominal inertia matrix; $\mathbf{C_n}(\mathbf{q},\dot{\mathbf{q}}) \in \mathbb{R}^{n \times n}$ represents the nominal Coriolis and centrifugal matrix; $\mathbf{g_n}(\mathbf{q}) \in \mathbb{R}^n$ represents the nominal gravity vector; $\boldsymbol{\tau} \in \mathbb{R}^n$ represents the joint torque vector; $\boldsymbol{\tau}_{\mathbf{dis}} \in \mathbb{R}^n$ represents the disturbance vector which includes plant uncertainties and external disturbances; and $\mathbf{q}, \dot{\mathbf{q}}$ and $\ddot{\mathbf{q}} \in \mathbb{R}^n$ represent the position, velocity and acceleration vectors of joints, respectively.

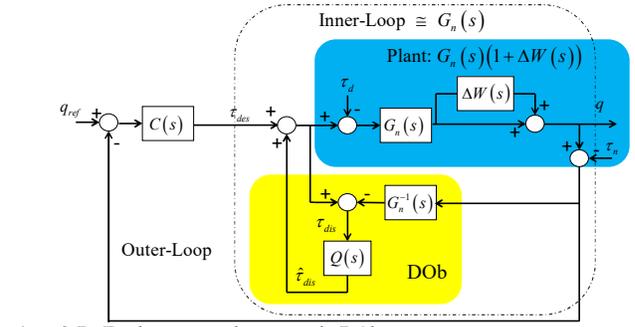

a) 2-DoF robust control system via DOb.

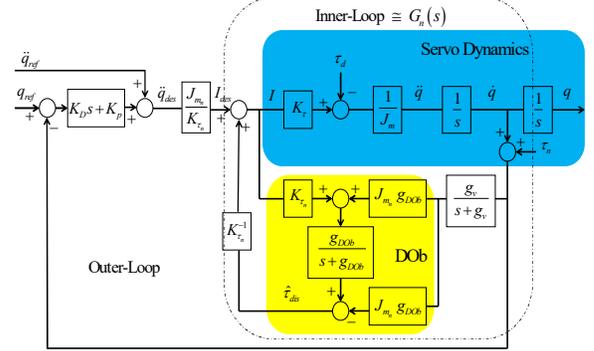

b) Acceleration-based robust position control system.

Fig.3: Block diagrams of 2-DoF robust control systems via DOb.

A DOb can be designed by using known nonlinear dynamic model of a robot manipulator as follows:

$$\hat{\boldsymbol{\tau}}_{\mathbf{dis}} = \mathbf{z} - L\dot{\mathbf{q}} \quad (20)$$

where $\hat{\boldsymbol{\tau}}_{\mathbf{dis}}$ is the estimation of $\boldsymbol{\tau}_{\mathbf{dis}}$; $L \in \mathbb{R}$ is an observer gain to be tuned; and $\mathbf{z}$ is derived by integrating

$$\dot{\mathbf{z}} = L\mathbf{M_n^{-1}}(\mathbf{q})\left(\boldsymbol{\tau} - \mathbf{C_n}(\mathbf{q},\dot{\mathbf{q}})\dot{\mathbf{q}} - \mathbf{g_n}(\mathbf{q}) - \hat{\boldsymbol{\tau}}_{\mathbf{dis}}\right) \quad (21)$$

The dynamic equation of disturbance estimation is derived by taking the derivative of Eq. (20) as follows:

$$\dot{\mathbf{e}}_{\boldsymbol{\tau}_{\mathbf{dis}}} = -L\mathbf{M_n^{-1}}(\mathbf{q})\mathbf{e}_{\boldsymbol{\tau}_{\mathbf{dis}}} - \dot{\boldsymbol{\tau}}_{\mathbf{dis}} \quad (22)$$

where $\mathbf{e}_{\boldsymbol{\tau}_{\mathbf{dis}}} = \hat{\boldsymbol{\tau}}_{\mathbf{dis}} - \boldsymbol{\tau}_{\mathbf{dis}} \in \mathbb{R}^n$; and $\frac{\partial \mathbf{L}(\mathbf{x})}{\partial \mathbf{x}} = L\mathbf{M_n^{-1}}(\mathbf{q})$.

If $L$ is strictly positive, then the stability (either asymptotic stability when $\dot{\boldsymbol{\tau}}_{\mathbf{dis}} = \mathbf{0}$ or uniformly ultimately bounded estimation error when $\boldsymbol{\tau}_{\mathbf{dis}}$ and $\dot{\boldsymbol{\tau}}_{\mathbf{dis}}$ are bounded) of disturbance estimation is achieved [78]. Eq. (22) shows that the stability and performance of disturbance estimation can be directly adjusted by tuning the nominal inertia matrix and the observer gain of DOb.

## V. 2-DoF ROBUST CONTROLLER SYNTHESIS

### A. Classical Control:

Block diagram of a DOb-based robust control system is illustrated in Fig. 3a. In this figure, $C(s)$ represents the performance controller, and $q_{ref}$ represents the exogenous reference input. The other parameters are same as defined earlier.

It has been experimentally verified in many applications that the performance controller can be tuned by considering only nominal plant dynamics in the outer-loop as DOb cancels plant uncertainties and external disturbances in the inner-loop.

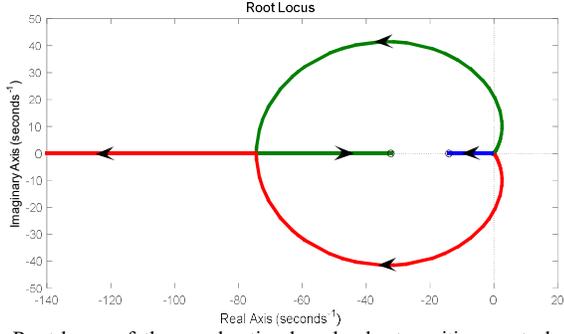

Fig.4: Root-locus of the acceleration-based robust position control system with respect to $\alpha$ when the LPF of velocity measurement is neglected.

Since the robustness and performance of the control system can be independently adjusted by tuning DOb and the performance controller in the inner and outer loops, respectively, Fig. 3a is widely known as a 2-DoF robust controller in the literature [24]. In fact, this assumption has some practical limitations. The inner-loop dynamics, i.e., imperfect disturbance estimation, may affect the stability and performance of the robust control system if the model-plant mismatches cannot be precisely compensated by DOb. It can be easily shown by deriving the transfer function between the reference input and output from Fig. 3a as follows:

$$\frac{q}{q_{ref}} = \frac{CG_n(s)(1+\Delta W(s))}{1+\Delta WQ(s)+CG_n(s)(1+\Delta W(s))} \quad (23)$$

where $\frac{q}{q_{ref}} \to \frac{CG_n(s)}{1+CG_n(s)}$ as $Q(s) \to 1$, i.e., $s = jw \to 0$ or the bandwidth of DOb goes to infinite.

Eq. (23) shows that the characteristic polynomial of the robust control system is influenced by the dynamics of the LPF of DOb, unstructured uncertainties, nominal plant model, and performance controller. The outer-loop transfer function of the robust control system can be free from the dynamics of the plant-model mismatches and the estimation of disturbances when the bandwidth of the LPF of DOb goes to infinite and/or the frequency of disturbances goes to zero. However, the former assumption is impractical as shown in Fig. 2 and the latter assumption is very strict for many engineering applications.

To show how the nominal plant dynamics influences the stability of the 2-DoF robust control system, let us consider the Acceleration-Based Controller (ABC) that is illustrated in Fig. 3b. In this figure, $K_P$ and $K_D$ represent the position and velocity control gains of the outer-loop performance controller, respectively. The other parameters are same as defined earlier.

When the dynamics of velocity measurement is neglected (i.e., $g_v \to \infty$), the transfer function between $\ddot{q}$ and $\ddot{q}_{des}$ is derived from Fig. 3b as follows:

$$\frac{\ddot{q}}{\ddot{q}_{des}} = \alpha \frac{s+g_{DOb}}{s+\alpha g_{DOb}} \quad (24)$$

Eq. (24) shows that a DOb can be designed as a phase lead-lag compensator by tuning $\alpha$ in the inner-loop. The phase margin, and thus the stability, of the robust position control system can be simply improved by increasing $\alpha$ as shown in Fig. 4. However, it has an upper bound in practice (See the waterbed effect due to increasing $\alpha$ in Fig. 2).

### B. Modern Control:

If a plant includes only matched disturbances which act through the same channel as that of the control input, then the 2-DoF robust controller can be similarly synthesized in state space. When a state feedback controller is used in the outer-loop, the dynamic model of the closed-loop system is derived by using Eq. (5) as follows:

$$\dot{\mathbf{x}} = (\mathbf{A_n} - \mathbf{b_n}\mathbf{K})\mathbf{x} + \mathbf{b_n}(\hat{\tau}_{dis} - \tau_{dis}) \quad (25)$$

where $\tau_{dis}$ and $\hat{\tau}_{dis} \in \mathbb{R}$ represent the matched disturbance and its estimation, respectively; and $\mathbf{K}^T \in \mathbb{R}^n$ represents the state feedback control gain.

The stability of the robust control system can be proved by using the following Lyapunov function candidate.

$$V = \mathbf{x}^T \mathbf{P} \mathbf{x} \quad (26)$$

where $(\mathbf{A_n} - \mathbf{B_n}\mathbf{K})^T \mathbf{P} + \mathbf{P}(\mathbf{A_n} - \mathbf{B_n}\mathbf{K}) = -\mathbf{Q}$; and $\mathbf{P}$ and $\mathbf{Q} \in \mathbb{R}^{n \times n}$ are positive definite matrices.

The derivative of Eq. (26) satisfies the following inequality.

$$\dot{V} \leq -\left(\min(\text{eig}(\mathbf{Q}))-1\right)\|\mathbf{x}\|^2 + \|\mathbf{Pb_n}(\hat{\tau}_{dis} - \tau_{dis})\|^2 \quad (27)$$

where $\min(\text{eig}(\mathbf{Q}))$ represents the slowest eigenvalue of $\mathbf{Q}$.

Eq. (27) shows that the time derivative of the Lyapunov function is negative outside of the compact set defined by

$$\Omega = \left\{\mathbf{x}(t) \in \mathbb{R}^n : \|\mathbf{x}(t)\|^2 \leq \frac{1}{\ell}\|\mathbf{Pb_n}(\hat{\tau}_{dis} - \tau_{dis})\|^2\right\} \quad (28)$$

where $\ell < \min(\text{eig}(\mathbf{Q})) - 1$.

Eq. (27) and Eq. (28) show that any states start out of the compact set ultimately enter in $\Omega$ when the outer-loop performance controller and DOb are properly tuned. As the accuracy of disturbance estimation improves, i.e., $\hat{\tau}_{dis} \to \tau_{dis}$, the bound of the compact set $\Omega$ shrinks. In addition to disturbance estimation, the stability of the overall robust control system can be improved by properly tuning the performance controller; e.g., the upper bound of the set $\Omega$ depends on $\mathbf{Q}$ and $\mathbf{P}$ as well as the disturbance estimation error as shown in Eq. (28).

If a plant includes not only matched but also mismatched disturbances which act through the different channels from that of the control input, then the robustness cannot be achieved by directly cancelling disturbances with their estimations via DOb. Different control techniques have been proposed to deal with the mismatched disturbances by using DOb [20, 114, 115]. In general, the state space model of the system can be reconstructed by using the estimations of mismatched disturbances and their successive time derivatives so that a new state space model which suffers from only matched disturbances is obtained [20, 114]. The reconstructed states of the system are bounded when DOb is properly tuned. Therefore, a stable robust controller can be similarly synthesized by using a state feedback controller and canceling the matched disturbances of the reconstructed state space model of the system with their estimations as shown in Eq. (25) [20]. Since the estimations of disturbances and their successive time derivatives are required to reconstruct the state space model with only matched disturbances, the 2-DoF robust controller becomes more sensitive to noise. The reader is recommended to refer to [20] for DOb-based robust control of a system with mismatched disturbances.

A 2-DoF robust controller can be similarly synthesized for a nonlinear system by implementing a DOb in the inner-loop and a nonlinear performance controller in the outer-loop [53, 62, 77, 78, 80, 81]. However, it is very hard to obtain a general 2-DoF robust control structure as different observers and nonlinear controllers are synthesized in the inner and outer loops, respectively [77, 78, 87]. The stabilities of the observer and 2-DoF robust control system can be analyzed by using Lyapunov's second method [77, 78].

## VI. CONCLUDING REMARKS

This paper has presented an overview on DOb-based robust control and its engineering applications. It is shown that the origins of DOb is as early as the origins of robust control theory which can be traced back to the end of 1960s. To improve the robustness of a Linear Quadratic Regulator, DOb-based robust control, for the first time, was proposed by using Gopinath's observer synthesis method in state space in 1983 [33]. However, DOb-based robust control has received significant attention, particularly from control engineering practitioners, with its frequency domain analysis and synthesis techniques. This has caused a common mistake in the literature. Many researchers have erroneously described DOb as a linear disturbance estimation method which is synthesized in Laplace domain. In fact, for the first time, a DOb was proposed by using auxiliary variable design method in time domain. Moreover, this design method is applicable for not only linear but also nonlinear systems as shown in Section IV.

Since a DOb-based robust controller can be intuitively synthesized without a strong mathematics background and easily implemented by using a simple microcontroller, control engineering practitioners have widely adopted this robust control technique and applied it to different engineering applications, particularly in the motion control and power electronic fields, in the last thirty-five years. Although a robust controller synthesis is the main driving force in the engineering applications of DOb, it has inspired many researchers to establish new practical control tools. For example, Murakami developed RFOb to estimate contact force by using DOb as a force sensor in [93], Natori developed Communication Disturbance Observer (CDOb) to estimate/compensate delay in network systems without using a model for delay in [65, 99], and Fujimoto developed Yaw-Moment Observer (YMOb) to improve steering stability and comfortability in electrical vehicles in [89].Today, commercial robust motion control products developed by using DOb are available in the market and show superior performance results over conventional PID controllers [95, 96]. This motivates researchers to develop new DOb-based practical control tools and commercial robust motion control products.

With the high performance motion control applications, many researchers from different fields have been attracted to the DOb-based robust control. Today, the examples of this robust control method can be found in Chemical Engineering [98], Telecommunications Engineering [99], Automotive Engineering [89], Aerospace Engineering [67], Biomedical Engineering [38], Renewable Energy Systems [116], Nuclear Science [105] and Biology [102], in addition to Electrical, Electronics and Mechanical Engineering [24, 111]. Several advanced control methods, such as nonlinear control, model predictive control, SMC and intelligent control, have been combined with DOb in order to apply the 2-DoF robust controller into complex systems. However, compared to motion control, DOb-based robust control and its applications are not matured in these fields yet. More practical and theoretical research should be conducted in order to improve the robust stability and performance of these applications, develop new analysis and synthesis techniques and establish new control tools such as RFOb.

The intuitive robust controller synthesis is one of the most important superiorities of DOb-based 2-DoF control over other robust control methods such as H∞ and μ-synthesis. Many successful robust control implementations of DOb have been reported by intuitively synthesizing the 2-DoF robust controller in the last three decades. The main drawback of this method is that the design parameters (i.e., the dynamics of the LPF of DOb, nominal plant model and outer-loop controller) are generally tuned by trial and error, so the performance of the robust controller highly depends on designers' own experience. To tackle this problem, several theoretical studies have been conducted in the last three decades. However, there is still lack of practical analysis and synthesis techniques for DOb-based robust control applications, particularly for systems with complex dynamics. The robust stability and performance of a DOb-based control system should be further investigated by considering the dynamics of the LPF of DOb (e.g., higher-order LPF) and nominal plant model (e.g., non-minimum phase plants) in addition to the bandwidth of DOb. For example, although one of the earliest applications of DOb is the robust motion control of a robot manipulator proposed in 1987, tuning the parameters of the nominal inertia matrix is still an open problem [35]. Experimental results show that the robust stability and performance of the motion controller may significantly deteriorate when the nominal inertia matrix changes [78]. The practical design constraints (e.g., sampling time in digital implementation and noise-sensitivity due to encoder reading in motion control) and conservatism, which may cause a severe limitation for the bandwidth of DOb, should be considered when new analysis and synthesis techniques are developed. Besides, more effort should be paid to understand the practical limitations, (e.g., bounds on the robustness and performance) of the DOb-based robust control systems.

Last but not least, developing standard analysis and synthesis tools for DOb-based control will help researchers easily adopt this robust control technique and apply it to various systems. To this end, H. Shim recently developed a new MATLAB toolbox for DOb-based robust control method, namely DO_DAT: *Disturbance Observer – Design & Analysis Toolbox*, [117]. Considering the significant increase in the popularity of DOb in the last decades, it is expected to see more examples of such control tools for DOb-based robust control method in the future.